\documentclass[hyper]{prop2015}
\usepackage[english]{babel}
\usepackage[all]{xy}
\usepackage{tikz-cd}

\usepackage{amsthm,bbm}
\theoremstyle{plain}
\newtheorem{theo}{Theorem}[section]

\newtheorem{propo}[theo]{Proposition}

\theoremstyle{definition}
\newtheorem{defi}[theo]{Definition}

\category{Proceedings}
\subtitle{\href{http://www.maths.dur.ac.uk/lms/109/index.html}{LMS/EPSRC Durham Symposium on Higher Structures in M-Theory}}
\title[Full Logarithmic Conformal Field Theory]
{Full Logarithmic Conformal Field theory --- an Attempt at a Status Report}
\author[J. Fuchs]{J\"urgen Fuchs\inst{a}}
\author[C.  Schweigert]{Christoph Schweigert\,\inst{b,}
\footnote{Corresponding author e-mail:~\href{mailto:christoph.schweigert@uni-hamburg.de}{\textsf{christoph.schweigert@uni-hamburg.de}}}}
\address[1]{Teoretisk fysik, \ Karlstads Universitet, Universitetsgatan 21, 65188 Karlstad, Sweden} 
\address[2]{Fachbereich Mathematik, \ Universit\"at Hamburg, Bundesstr. 55, 20146 Hamburg, Germany}
\shortauthors{J. Fuchs and C. Schweigert}
\begin{acknowledgements}
JF is supported by VR under project no.\ 2017-03836. CS is partially supported by the 
RTG 1670 `Mathematics inspired by String theory and Quantum Field Theory'
and by the Deutsche Forschungsgemeinschaft (DFG, German Research Foundation) under
Germany's Excellence Strategy - EXC 2121 `Quantum Universe'- QT.2.
\end{acknowledgements}

\begin{abstract}
Logarithmic conformal field theories are based on vertex algebras with non-semisimple 
representation categories. While examples of such theories have been known for more than 25 years, 
some crucial aspects of local logarithmic CFTs have been understood only recently, with the help of
a description of conformal blocks by modular functors. We present some of these results, 
both about bulk fields and about boundary fields and boundary states. 
We also describe some recent progress towards a derived modular functor. 
\\
This is a summary of 
work with Terry Gannon, Simon Lentner, Svea Mie\-rach, Gregor Schaumann and Yorck Sommerh\"auser.
\end{abstract}
\shortabstract
\begin{document}

\maketitle
\def\Bl     {\mathrm{Bl}}
\def\C      {{\mathcal C}}
\def\D      {{\mathcal D}}
\def\FM     {{\mathcal F\!\!\!\!\mathcal M}}
\newcommand{\func}[2]{#1 \left( #2 \right)}
\def\Hom    {\mathrm{Hom}}
\def\mSurfc {m\Surfc}
\def\Surfc  {\mathcal S\!\text{\it urf}}

\section{Introduction} \label{seccs:1}

Let us start this report on logarithmic conformal field theories
by explaining the qualification `logarithmic'.
To this end we first recall textbook knowledge about ordinary two-dimensional
conformal field theories. Consider a Virasoro primary field $\phi(z)$ of 
conformal weight $h$. Via the state-field correspondence, it gives an eigenstate 
$|\phi\rangle$ of the Virasoro zero mode $L_0$ with eigenvalue
$h$, i.e.\ $L_0|\phi\rangle \,{=}\, h\, |\phi\rangle$. The operator product
between the stress-energy tensor $T$ and such a chiral field $\phi$ takes the form
  \begin{equation}
  T(z)\, \phi(w) \,\sim\, \frac{h\, \phi(w)} {(z-w)^2} + \frac{\partial \phi(w)} {z-w} \,.
  \end{equation}
This amounts to the commutation relations 
  \begin{equation}
  \begin{aligned}
  [L_{-1},\phi(w)] &= \partial\phi(w) \,, \\
  [L_0,\func{\phi}{w}] &= h\, \func{\phi}{w} + w\, \partial \func{\phi}{w} \,,\\
  [L_1,\func{\phi}{w}] &= 2h\, w\, \func{\phi}{w} + w^2 \partial \func{\phi}{w}
  \end{aligned}
  \end{equation}
with the Laurent modes $L_n$ of the stress-energy tensor.

Combining these relations with the invariance of the vacuum state under 
the Lie algebra $\mathfrak{sl}(2,\mathbb{C})$ 
that is spanned by the modes $L_0$, $L_1$ and $L_{-1}$ leads to
the following differential equations for the two-point conformal blocks
of the field $\phi$:
  \begin{equation}
  \begin{aligned}
   (\partial_z + \partial_w) \, \langle\func{\phi}{z}\, \func{\phi}{w}\rangle &= 0 \,, \\
   (z\, \partial_z + w\, \partial_w + 2h) \, \langle\func{\phi}{z}\, \func{\phi}{w}\rangle &= 0 \,,\\
   (z^2 \partial_z + w^2 \partial_w + 2h\, (z\,{+}\,w)) \,  \langle\func{\phi}{z}\, \func{\phi}{w}\rangle &= 0 \,.
  \end{aligned}
  \end{equation}
The general solution of these equations exhibits scaling behaviour:
  \begin{equation}
  \langle\phi(z) \phi(w)\rangle = \frac{A}{(z-w)^{2h}}
  \end{equation}
for some constant $A$.

In a logarithmic conformal field theory, the action of $L_0$ need not be
semisimple, so that a Jordan partner $|\Phi\rangle$ of $|\phi\rangle$ can appear, satisfying
$L_0 |\Phi\rangle = h |\Phi\rangle + |\phi\rangle$ in addition to 
$L_0 |\phi\rangle = h |\phi\rangle$. This amounts to the operator products
\begin{subequations}
  \begin{equation}
   \func{T}{z} \func{\Phi}{w} \,\sim\, \frac{h\, \func{\Phi}{w}
  + \func{\phi}{w}}{(z-w)^2} + \frac{\partial \func{\Phi}{w}}{z-w} \
  \end{equation}
  and
  \begin{equation}
  T(z)\, \phi(w) \,\sim\, \frac{h\, \phi(w)} {(z-w)^2} + \frac{\partial \phi(w)} {z-w} 
  \end{equation}
  \end{subequations}
with the the stress-energy tensor.
Accordingly, the Virasoro modes $L_0$ and $L_{\pm1}$ act on the Jordan partner $\Phi$ as
   \begin{equation}
  \begin{aligned}
  [L_{-1},\func{\Phi}{w}] &= \partial \func{\Phi}{w} \,, \\
 [L_0,\func{\Phi}{w}] &= h\, \func{\Phi}{w} + w\, \partial \func{\Phi}{w}
  + \func{\phi}{w} \,,\\
   [L_1,\func{\Phi}{w}] &= 2h\, w\ \func{\Phi}{w} + w^2 \partial
  \func{\Phi}{w} + 2 w\, \func{\phi}{w} \,.
  \end{aligned}
  \end{equation}

This leads to the a set of inhomogeneous differential equations for the 
conformal blocks. For the full set of these equations and for an extensive exposition of
vertex algebras leading to logarithmic conformal field theories, see e.g.\ \cite{Creutzig:2013hma}.

The equations for the two-point blocks include in particular
\begin{subequations}
  \begin{equation}
  (\partial_z + \partial_w) \, \langle\func{\phi}{z}\, \func{\Phi}{w}\rangle = 0 
  \end{equation}
and
  \begin{equation}
  (z\, \partial_z + w\, \partial_w + 2h) \, \langle\func{\phi}{z}\, \func{\Phi}{w}\rangle
  =-\, \langle\func{\phi}{z}\, \func{\phi}{w}\rangle \,.
  \end{equation}
  \end{subequations}

Let us simplify the discussion by assuming that the two fields $\func{\phi}{z}$
and $\func{\Phi}{z}$ are mutually bosonic, i.e.\ that
$\langle\func{\phi}{z}\,\func{\Phi}{w}\rangle \,{=}\, \langle\func{\Phi}{w}\,\func{\phi}{z}\rangle$.
In this case the two-point blocks take the form
     \begin{equation}
  \begin{aligned}
 \langle\func{\phi}{z}\, \func{\phi}{w}\rangle &= 0 \,,
  \\
\langle\func{\phi}{z}\, \func{\Phi}{w}\rangle &= \frac{B}{(z-w)^{2h}}~,
  \\
\langle\func{\Phi}{z}\, \func{\Phi}{w}\rangle &= \frac{C - 2B\, \log\, (z-w)}{(z-w)^{2h}} 
  \end{aligned}
  \end{equation}
for some constants $B$ and $C$. Thus when the $L_0$-action is non-diagonalizable,
imposing global conformal invariance gives rise to the presence of
logarithmic singularities in conformal blocks.

The rest of this contribution will not involve any of those logarithms. The crucial
point is rather that we will \emph{not} require semisimplicity, thereby allowing
for the occurrence of Jordan blocks in the action of the chiral algebra. Accordingly, from
now on we prefer to talk of \emph{non-se\-misimple conformal field theory}, rather than of
logarithmic CFT. More specifically, we will work with monoidal categories 
that, unlike in rational conformal field theory, are not required to be semisimple.
Readers not fully conversant with the theory of monoidal categories should
feel free to think of these categories as realized concretely by representations of 
suitable vertex algebras and by intertwiners between such representations, with the 
tensor product given by fusion.

At this point it is appropriate to point out that the past decade has
seen a lot of progress in the understanding of specific classes
of conformal vertex algebras that have such representation categories;
see e.g.\ \cite{Fjelstad:2002ei,Miyamoto:0209101,Feigin:2005zx,Feigin:2005xs,Adamovic:2007er,Huang:2007mj,Tsuchiya:2012ru,Arakawa:2016yja,Creutzig:2014nua,Lentner:2017dkg,Farsad:2017eef,Creutzig:2017gpa} 
for a biased selection of references. Under very general conditions, 
a conformal vertex algebra is expected to possess a representation category that is endowed
with a braided monoidal structure. Moreover, the braiding is expected to be non-degenerate, 
in a sense to be made precise below. 

We further restrict ourselves to vertex algebras with a representation category that obeys certain finiteness
conditions; technically, it is required to be a \emph{finite tensor category} in the 
sense of \cite[Ch.\,6]{Etingof:2015aa}. Finally, we require that the category comes with
dualities and with a compatible balancing (or twist), whereby it acquires the structure
of a \emph{ribbon} category. Altogether this means that we work with a category $\mathcal C$
that is a \emph{factorizable} finite ribbon category, or \emph{modular} tensor category 
in the terminology of \cite{Shimizu:1602.06534} (for further
explanations, see below). In recent years much progress has been made in the understanding
of modular tensor categories and in the construction of examples.
An attractive feature of some of those examples is that they
are directly connected with Lie-theoretic structures, whereby they promise
to yield models that are amenable to detailed explicit computations
(as illustrative examples, see e.g.\ \cite{Flandoli:2017xwu,Lentner:2016hex}, which reflect recent PhD work).

It is worth pointing out that the structures mentioned so far are all related to \emph{chiral}
conformal field theory. A priori, given the fact that in non-semisimple theories the 
conformal blocks can contain logarithms, it is by no means clear whether
a \emph{full}, local conformal field theory, with correlators that are single-valued functions, 
can be constructed from the conformal blocks of a chiral logarithmic conformal field theory.
First encouraging results that suggest that this is nevertheless possible date back 
almost two decades (see \cite{Gaberdiel:1998ps}, as well as \cite{Gaberdiel:2010rg} and references therein for more 
recent work). However, a systematic construction of local non-semisimple
conformal field theories has been elusive for a long time.
Indeed, not even the existence of the local conformal field theory that generalizes the
charge-conjugate partition function of a rational CFT could be established.
(Incidentally, even in the rational case the consistency of that full CFT was fully
established only relatively recently \cite{Felder:1999mq}.
In this context it may also be of interest that there exist chiral rational CFTs to which
there isn't associated a consistent full CFT with a torus partition function given by the
`true diagonal' modular invariant \cite{Sousa:2002te}.)

A thorough understanding of such field theories is highly desirable. After all,
local non-semisimple conformal field theories do have significant applications in the real
world, for instance to critical dense polymers \cite{Duplantier:Exact} or to critical percolation
\cite{Cardy:1991cm}. (Applications to string theory seem to be more speculative at the
time of writing; see, however \cite{Witten:2018xfj}.)


\section{Strategy}

Our goal is thus to find a model-independent construction of local non-semisimple
conformal field theories from a given chiral theory.
Our general strategy to address this issue is as follows. First recall
how conformal blocks can be obtained for a vertex algebra, see \cite{Sorger:La}
for a review in the case of WZW theories, and \cite{Frenkel:2004qq} for a more general
approach. The $n$-point conformal blocks on a genus-$g$ surface form 
the total space of a vector bundle $\mathcal V$ with projectively flat connection
over the moduli space ${\mathcal M}_{g,n}$ of complex curves of genus $g$ with $n$ 
local holomorphic coordinates (which we will later tacitly replace by $n$ disjoint 
boundary circles). Given an $n$-tuple $\big({\cal H}_{\lambda_i}^{}\big)_{i=1,\ldots,n}^{}$
of modules over a vertex algebra -- that is, an $n$-tuple of objects of the
representation category $\mathcal C$, which is a modular tensor category -- the bundle
$\mathcal V \,{=}\, \mathcal V_{\lambda_1,\ldots,\lambda_n}$ is concretely realized
as a subbundle of invariants under the action of a globalized version of the vertex algebra,
  \begin{equation}
  \begin{tikzcd}[row sep=2.8em]
  {\mathcal M}_{g,n}\times\left({\cal H}_{\lambda_1}{\otimes}\,\cdots\,{\otimes}
  {\cal H}_{\lambda_n}\right)^*_{}  \ar{d}  \ar[hookleftarrow]{r}
  & {\mathcal V}_{\lambda_1,\ldots,\lambda_n} \ar{dl}
  \\
  {\mathcal M}_{g,n}
  \end{tikzcd}
  \end{equation}
In general, horizontal sections of this bundle are mul\-ti-va\-lued, i.e.\ they
exhibit non-trivial monodromies under analytic continuation. Hence only very specific sections
-- very specific conformal blocks -- can qualify as correlators of a full CFT. 

Monodromies for sections of $\mathcal V$ are, after choosing a base point, 
encoded in representations of mapping class groups
$\pi_1({\mathcal M}_{g,n}) \,{=}\, \mathrm{Map}_{g,n}$.
These representations are required to depend functorially on the objects ${\cal H}_{\lambda_i}$
in the modular tensor category $\mathcal C$.
A tool for keeping keep track of these representations and their functorial
dependence on objects in $\mathcal C$ is provided by a \emph{modular functor}.
The latter is a symmetric monoidal 2-functor from a suitable bicategory of bordisms to
an algebraic bicategory. For the bordisms we take the bicategory $\mathrm{Bord}_{2,1}$; objects
of $\mathrm{Bord}_{2,1}$ are closed oriented one-dimensional manifolds, 1-morphisms are surfaces 
with parametrized boundaries, and 2-morphisms are given by elements of mapping class groups
(these will be explicitly described further below); the tensor product is disjoint union.
 
We obtain a specific tractable framework by imposing the following finiteness condition on the 
target bicategory of the modular functor: We take the symmetric monoidal bicategory of 
\emph{finite tensor
categories}, with left exact functors as 1-morphisms, natural transformations as 2-morphisms,
and the Deligne product $\boxtimes$ of finite abelian categories as the tensor product.
Thus we give

\begin{defi}
A {\it modular functor} is a symmetric monoi\-dal 2-functor from
the bicategory $\mathrm{Bord}_{2,1}$ to the bicategory of finite tensor categories.
\end{defi}

Such a modular functor can indeed be constructed when $\mathcal C$ is taken to be any modular
tensor category; thus semisimplicity needs not to be imposed, provided that the finiteness
conditions are kept. Still, we hope that some of our structural results will extend beyond 
this class of categories, thereby also covering models like e.g.\ Liouville theory.
In any case, the study of non-se\-mi\-simple conformal field theories forces us to use
rather systematically concepts and tools from category theory. Thereby it allows for
a substantial conceptual clarification which, in turn, further elucidates also the structure 
of (semisimple) rational conformal field theories.

The rest of these notes is organized as follows. 
In order to construct a modular functor we develop, in Section \ref{seccs:2},
a Lego-Teichm\"uller game for a factorizable finite ribbon category $\D$
that is not necessarily semisimple. This is based on \cite{Fuchs:2016wjr},
which combines earlier work on the semisimple case \cite{Bakalov:9809057} with
categorical constructions introduced in \cite{Majid:Braided,Lyubashenko:1994tm,Lyubashenko:Ribbon,Kerler:1995jw}. For applications to
bulk fields in full conformal field theory, the category $\D$ has the form of a Deligne product
  \begin{equation}
  \D = \C\boxtimes \C^\text{rev} \,, 
  \end{equation}
where $\C$ is a modular tensor category and $\C^\text{rev}$ is the same rigid 
monoidal category as $\C$ but with reversed braiding and twist. 

In Section \ref{seccs:3} we discuss
correlators of bulk fields. We introduce the notion of a consistent system of bulk correlators
for a modular tensor category $\D$ and show that such systems are in bijection with modular
Frobenius algebras in $\D$. In Section \ref{seccs:4b} we present specific results for the torus
partition function. The next two sections contain complementary
material: Section \ref{seccs:4} deals with boundary states of non-semisimple 
conformal field theories; in Section \ref{seccs:5} we present some first results
towards a derived modular functor. In the final Section \ref{seccs:6}
we collect a few open problems and conceptual questions.


\section{A Lego-Teichm\"uller game with coends} \label{seccs:2}

\subsection{Preparations}

We now present an explicit construction of a modular functor with input datum a not
necessarily semisimple modular tensor category $\D$. A closed oriented one-ma\-ni\-fold -- an 
object of $\mathrm{Bord}_{2,1}$ -- is a disjoint union of finitely many, say $n$, copies of
an oriented circle ${\mathbb S}^1$. We assign to such a manifold the $n$-fold Deligne product
$\D^{\boxtimes n}$. To tackle 2-manifolds and their mapping class groups, we 
set up a \emph{Lego-Teichm\"uller game}, which associates (left exact) functors to surfaces.
To this end, we need to specify the appropriate classes of surfaces.

\begin{defi}
\begin{enumerate}[i)]
\item
An {\it extended surface} $(E,\partial_\text{in}E,\partial_\text{out}E,\linebreak
\{p_\alpha\})$ consists of a smooth oriented surface $E$ with oriented boundary.
The set of boundary components is partitioned into incoming and outgoing boundaries,
$\partial E \,{=}\, \partial_\text{in}E \,{\cup}\, \partial_\text{out}E$. 
On each boundary component $\alpha$ a marked point $\{p_\alpha\}$ is chosen
(this rigidifies the situation and will allow for a convenient description of
Dehn twists around boundary circles).

\item
The {\it mapping class group} $\mathrm{Map}(E)$ is the group of isotopy classes 
of orientation preserving diffeomorphisms from $E$ to itself that on the boundary
restrict to maps
$\partial_\text{in}E\,{\to}\, \partial_\text{in}E$ and $\partial_\text{out}E
\,{\to}\, \partial_\text{out}E$ and that map marked points to marked points.
\end{enumerate}
\end{defi}

We also need to introduce the additional operation of a \emph{sewing} of surfaces.
This produces a new extended surface from an existing one: select a pair $(\alpha,\beta)$
consisting of an incoming and an outgoing boundary component of $E$ and obtain a
new extended surface $\bigcup_{\alpha,\beta}E$ by identifying the circles $\alpha$
and $\beta$ (including their marked points). The following picture illustrates how
this procedure can be performed iteratively to sew three spheres having two, three and six holes,
respectively, to a torus with five holes:
  \begin{equation}
  \begin{picture}(220,225)(0,0)
  \put(-8,108) {\scalebox{.27} {\includegraphics{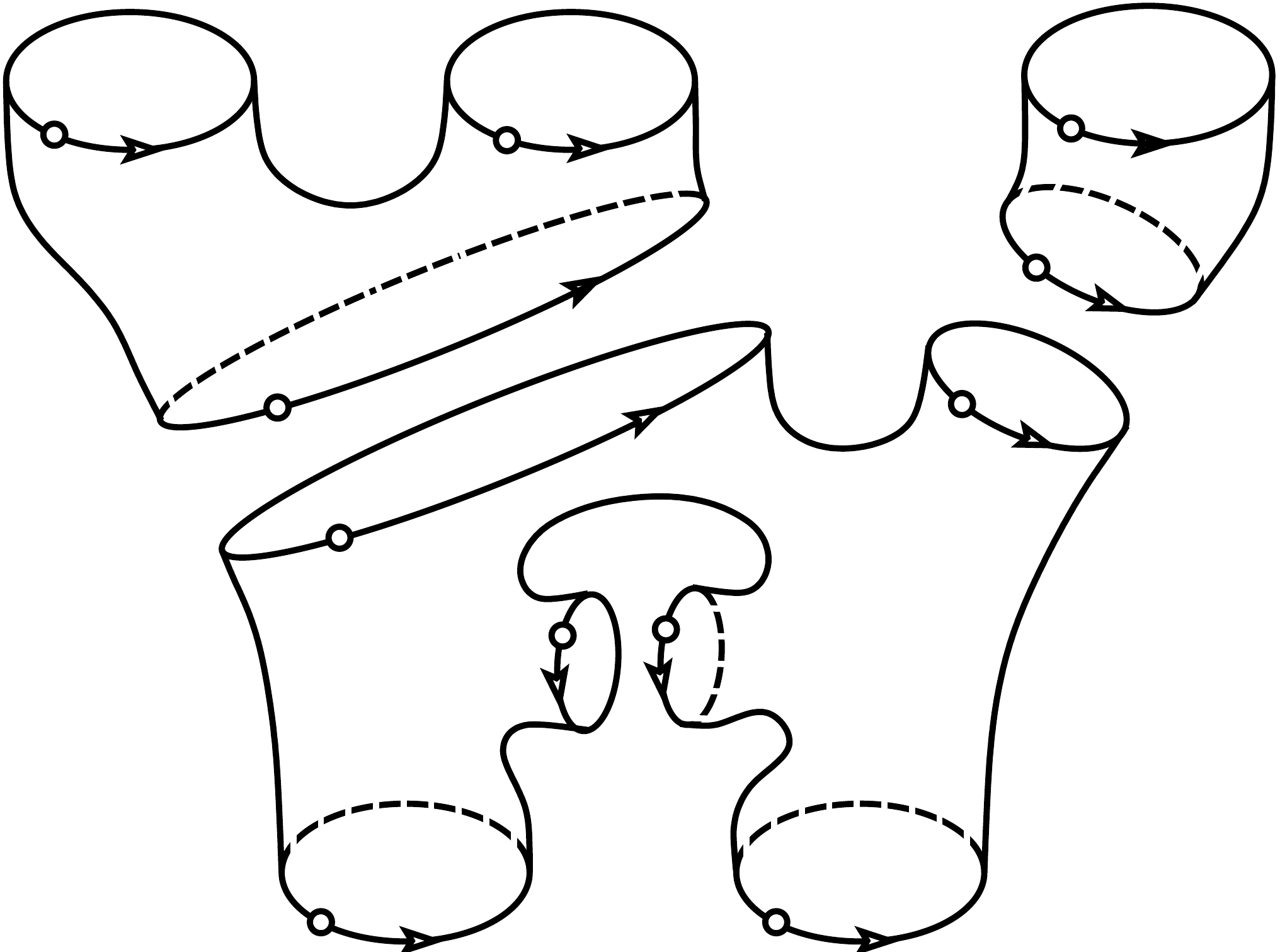}} }
  \put(37,32) {\large{\boldmath{$ \xrightarrow{~~~\raisebox{4pt}{sew}~~~} $}}}
  \put(78,-4) {\scalebox{.27} {\includegraphics{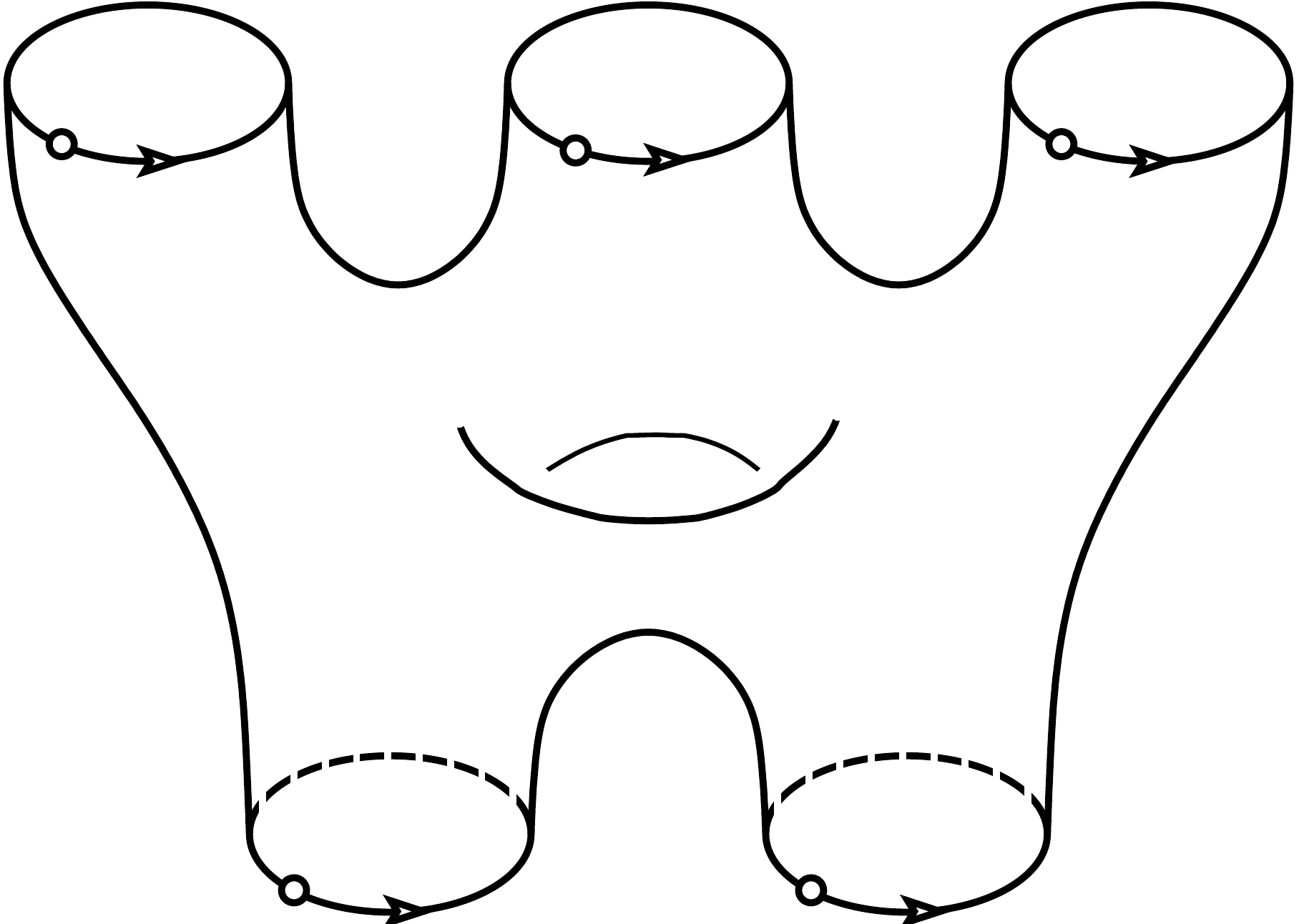}} }
  \end{picture}
  \end{equation}

To be able to construct a modular functor, we need to provide auxiliary
structure on the surface $E$, which in particular specifies a 
pair-of-pants decomposition of $E$:

\begin{defi}
\begin{enumerate}[i)]
\item
A {\it cut system} for $E$ is a finite set of disjoint oriented circles on $E$, each with a
marked point, that induces a decomposition of $E$ into punctured spheres. 
\\
A cut system is called {\it fine} iff each sphere in this decomposition has at most three 
punctures; the corresponding decomposition is then called a pair-of-pants decomposition of $E$.

\item
A {\it fine marking} on $E$ is a fine cut system together with a graph $\varGamma$ on $E$ 
with one vertex $q_{\!j}$ in the interior of each punctured sphere $\mathbb P_{\!\!j}$ of the
pair-of-pants decomposition that results from the cut system and, for each $j$, with an
edge pointing from $q_{\!j}$ to each of the marked points on the boundary of $\mathbb P_{\!\!j}$
(which consists of boundary circles of $E$ and/or cuts). 
\\
As an additional structure, for each of the graphs $\varGamma_{\!\!\!\!j}$ that are obtained by
restricting $\varGamma$ to the spheres $\mathbb P_{\!\!j}$, one edge is considered as
distinguished. This choice of distinguished edge induces a linear order 
(refining the cyclic order provided by the orientation of $\mathbb P_{\!\!j}$)
on the set of edges of the graph $\varGamma_{\!\!\!\!j}$.
\end{enumerate}
\end{defi}

As an illustration of these concepts, consider the five-ho\-led torus shown above.
The following picture shows this surface together with a fine cut system on it, for 
which the resulting pair-of-pants decomposition is the disjoint union 
of one two-holed sphere and five three-holed spheres
(for better readability, the 1-orientation of the cuts is suppressed):
  \begin{equation}
  \begin{picture}(240,114)(0,0)
  \put(47,0)   {\scalebox{.29} {\includegraphics{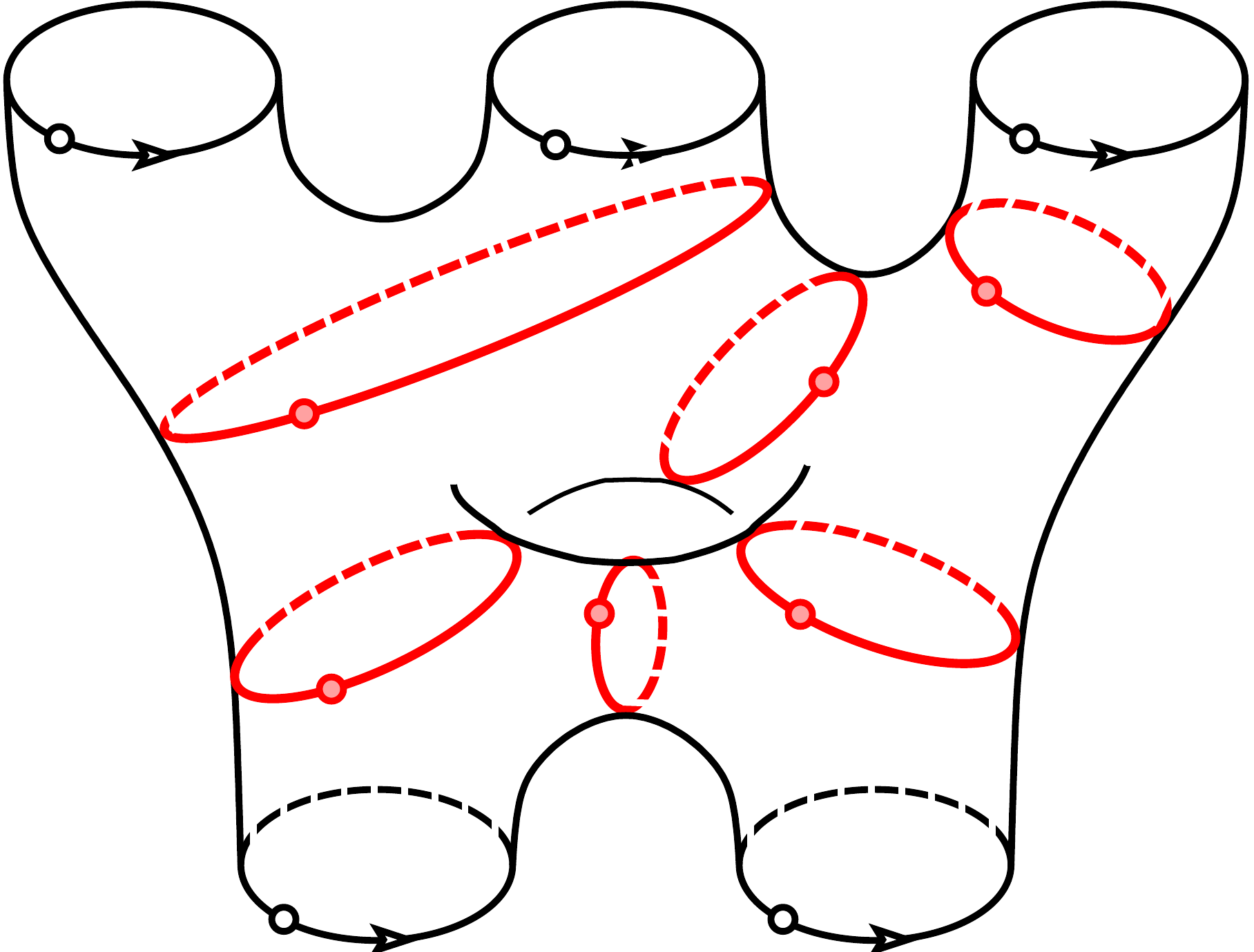}} }
  \end{picture}
  \end{equation}
A fine marking for this cut system is displayed in the next picture;
the distinguished edges of the subgraphs on the five spheres of the
pair-of-pants decomposition are accentuated by a small triangular flag:
  \begin{equation}
  \begin{picture}(240,115)(0,0)
  \put(47,3)  {\scalebox{.29} {\includegraphics{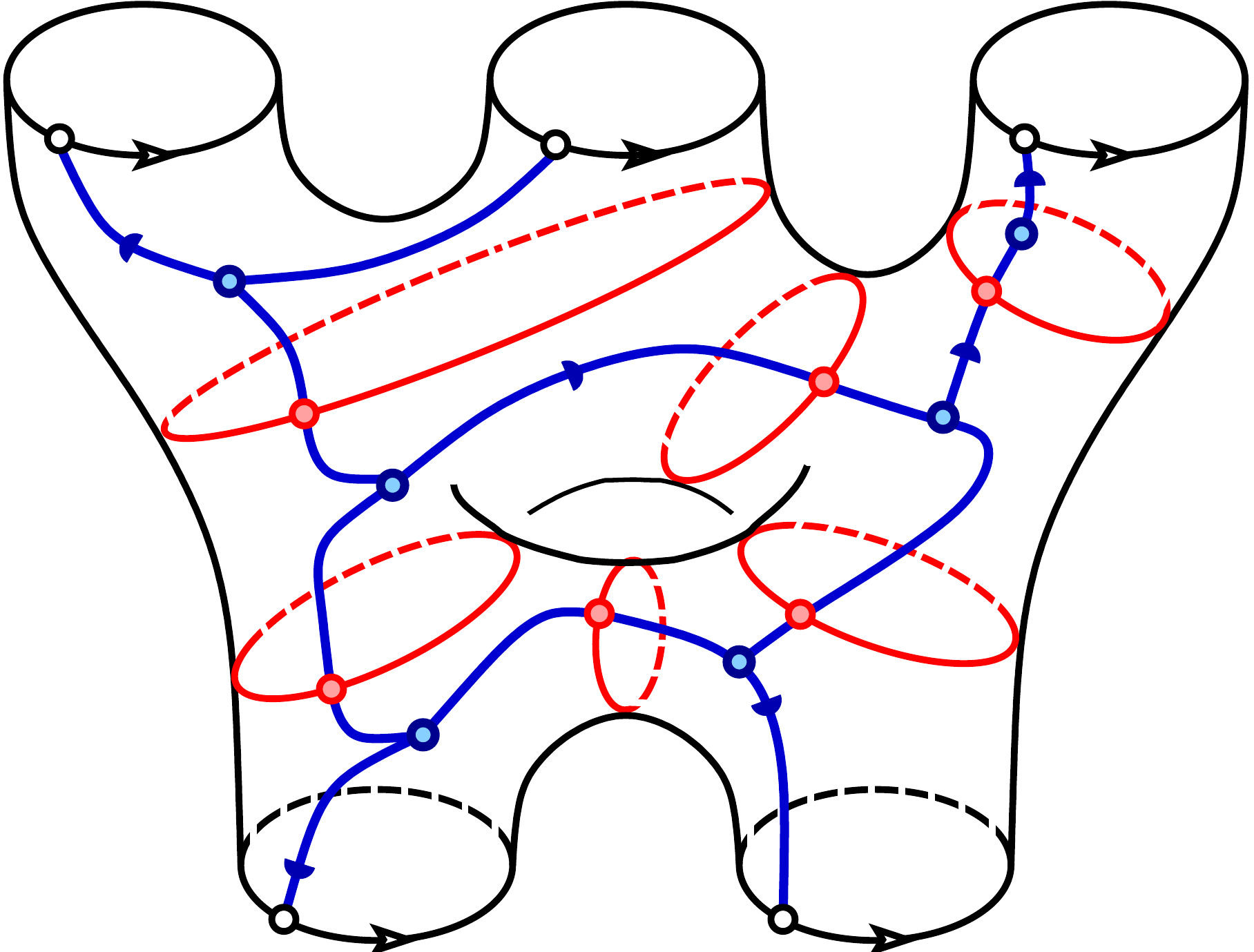}} }
  \end{picture}
  \end{equation}
		
Given any extended surface $E$, one can set up a group\-o\-id $\FM(E)$
of fine markings on $E$ \cite{Bakalov:9809057}. The objects of $\FM(E)$ are fine markings 
$(E,\varGamma)$ of $E$ (we suppress the cut system in the notation), and its morphisms 
are sequences of elementary \emph{moves}
$(E,\varGamma) \,{\mapsto}\, (E,\varGamma')$, modulo relations among the elementary moves. In more detail, there are 5 elementary moves:

\smallskip
\begin{enumerate}[(M1)]
\item
The \emph{$Z$-move}, which changes the distinguished edge of the graph on a
two- or three-holed sphere (without cuts) cyclically \cite[Figure\,8]{Bakalov:9809057}.
\item
The \emph{$B$-move}, which changes the graph on a three-holed sphere (without cuts)
in the same way as a certain braiding diffeomorphism \cite[Figure\,10]{Bakalov:9809057}.
\item
The \emph{$F$-move}, which implements `fusion' by removing a single cut from
a marking on a three-holed sphere \cite[Figure\,9]{Bakalov:9809057}.
\item
The \emph{$A$-move}, which implements `associativity' for different pair-of-pants
decompositions of a four-punc\-tu\-red sphere, by replacing a single cut on the sphere by
another, non-isotopic, cut \cite[Figure\,20]{Bakalov:9809057}.
\item
The \emph{$S$-move}, which implements the exchange of the two cycles 
in a symplectic homology basis for a ge\-nus-one surface \cite[Figure\,16]{Bakalov:9809057}.
\end{enumerate}
\smallskip

\noindent
Among these elementary moves there are 13 types of relations. We refrain from giving
a complete list (they can e.g.\ be found in \cite[Sect.\,2.2]{Fuchs:2016wjr}). Suffice
it to say that among them there are a pentagon relation for the $A$-mo\-ve, a hexagon 
relation involving the $A$- and $B$-moves, and $\mathrm{SL}_2(\mathbb{Z})$-relations for the 
one-punctured torus.

The following results about surfaces with markings are important to us:

\smallskip
\begin{enumerate}[i)]
\item
For any extended surface $E$, the groupoid $\FM(E)$ of fine markings is a 
connected tree \cite{Bakalov:9809057}. 
\\
Concretely, one can pass from any fine marking on $E$ to any other fine marking by a sequence of
elementary moves that is unique up to known relations.

\item
There is an \emph{unmarking functor}
  \begin{equation}
  U:\quad \FM(E)\stackrel\simeq\longrightarrow E/\!/ \mathrm{Map}(E) \,.
  \end{equation}
Here $E/\!/\mathrm{Map}(E)$ is the one-object groupoid with object
$E$ and endomorphisms given by the mapping class group of $E$.
On objects, the functor $U$ forgets the marking. On morphisms, it is determined
by sending the $F$-move to the identity morphism and each of the other elementary
moves to the uniquely determined mapping class that has the same
effect on the marking of $E$ as that move. The functor $U$ is an equivalence of groupoids.
\end{enumerate}
\smallskip

As a further input we need the following categorical structure, which has been known for more
than two decades \cite{Majid:Braided,Lyubashenko:1994tm}: for a finite ribbon category $\D$ the coend
  \begin{equation}
  K:=\int^{X\in\D}\! X^\vee\otimes X
  \end{equation}
has a natural structure of a Hopf algebra internal to $\D$ and comes with a Hopf pairing
$\,\omega\colon K\,{\otimes}\, K \,{\to} {\mathbf1}$ and with an integral and cointegral. 

It turns out that there is an intimate relation between this structure and the
\emph{Drinfeld center} ${\mathcal Z}(\D)$ of $\D$, i.e.\ the category whose objects
are pairs of an object $X$ of $\D$ and a half-braiding on $X$. The Drinfeld center
of any monoidal category is braided. If the category $\D$ itself is already brai\-ded
(as is the case in the situation at hand), the braiding and opposite braiding on $\D$
give rise to two braided functors $\D \,{\to}\, {\mathcal Z}(\D)$ and
$\D^\text{rev}\,{\to}\, {\mathcal Z}(\D)$; these functors combine into a braided functor
  \begin{equation}
  G_\D: \quad \D\boxtimes \D^\text{rev}\xrightarrow{~~~} {\mathcal Z}(\D) \,.
  \end{equation}
As has been shown relatively recently \cite{Shimizu:1602.06534}, the functor $G_\D$ is a braided
equivalence if and only if the Hopf pairing $\omega$ on the Hopf algebra $K \,{\in}\, \D$
is non-degenerate. A finite ribbon category obeying this non-degeneracy condition
for the braiding is called a \emph{modular tensor category}. This reduces to the
traditional notion of modular tensor category in the case that $\D$ is finitely semisimple. 

The coend $K$ contains in fact a lot of relevant information. In particular, if $\D$ 
is finitely semisimple and modular, then the Reshetikhin--Turaev invariants for closed
oriented  three-manifolds can be expressed in terms of the Hopf algebra $K$ and 
its integral only (see e.g.\ \cite[Corollary\,3.9]{Virelizier:0312337}).

We will make use of several facts about coends (for a short summary of pertinent
information see e.g.\ \cite{Fuchs:2016whb}). 
The two most important features for us are:

\smallskip
\begin{enumerate}[i)]
\item
There is a Fubini theorem for coends: in multiple coends the order can be interchanged.

\item
Coends in categories of left exact functors between finite tensor categories are
representable. 
\\
Specifically, we have e.g.
  \begin{equation}
  \int^{X\in\D} \! \Hom_\D(\reflectbox?\,,X) \otimes \Hom_\D(X\,,?) = \Hom_\D(\reflectbox?\,,?)
  \end{equation}
(which is a special instance of the Yoneda lemma) and
  \begin{equation}
  \int^{X\in\D} \! \Hom_\D(\reflectbox?\,,?\otimes X^\vee {\otimes}\, X)
  = \Hom_\D(\reflectbox?\,,?\otimes K)
  \end{equation}
with $K \,{=}\, \int^{X\in\D}\! X^\vee{\otimes}\, X$ as above.
\end{enumerate}


\subsection{A Lego-Teichm\"uller game} \label{sec32}

We are now ready to construct the left exact functors that are needed for a
modular functor via a Lego-Teichm\"uller game based on a, not necessarily semisimple,
modular tensor category $\D$. We proceed in two steps.

In the first step we associate a left exact functor 
  \begin{equation}
  \widetilde\Bl: \quad
  \FM(E) \xrightarrow{~~~} \mathcal{L}\!\text{\it ex}(\D^{\boxtimes p},\D^{\boxtimes q})
  \end{equation}
to an extended surface $E$ with fine marking for which $\partial E$ consists of
$p$ incoming and $q$ outgoing circles. This step combines two principles:

\smallskip
\begin{enumerate}[i)]
\item
For surfaces of genus zero, one implements the fact that blocks can be realized as (co-)invariants
(with respect to the globalized action of a vertex operator algebra, say) by taking
$\widetilde\Bl$ to be an appropriate Hom functor.
   
\item
Sewing is realized via the idea of `summing over all intermediate states' which
(as realized in \cite{Lyubashenko:1994tm}) is concretely implemented by taking coends of left exact functors.
\end{enumerate}

\smallskip

Thus we start with a sphere with at most three holes, such as $E\,{=}\,\mathbb{S}_{0|3}$
(i.e.\ a sphere with three outgoing and without incoming punctures) with some marking
$\varGamma$. We assign to it the left exact functor given by
  \begin{equation}
  \widetilde\Bl_{\mathbb{S}_{0|3}^{},\varGamma} :\quad
  X_1\boxtimes X_2\boxtimes X_3\, \xmapsto{~~~} \, \Hom_\D(1,X_1\,{\otimes}\,X_2\,{\otimes}\,X_3)
  \end{equation}
for $X_1,X_2,X_3 \,{\in}\, \D$, where the order of tensor factors is determined by 
the graph $\varGamma$. This prescription combines the first principle
with the idea that at genus zero the multi-point situation should amount to a tensor product.

To get the functors for more general surfaces $E$ we perform multiple sewings, with the precise
form of the corresponding coends prescribed by the cut system of the chosen marking on $E$.
By the Fubini theorem the order of the sewings is irrelevant, so that we obtain
a left exact functor 
  \begin{equation}
  \widetilde\Bl_{E,\varGamma}(-)
  = \int^{Y_1\boxtimes\cdots\boxtimes Y_\ell
  }\!\! \Big( \bigotimes_{i=1}^s \widetilde\Bl_{\mathbb P_{\!i},\varGamma_{\!\!\!i}} \Big)
  (-\,,Y_{\!1}^{},Y_{\!1}^\vee\!,\ldots\,,Y_{\!\ell}^{},Y_{\!\ell}^\vee) \,,
  \end{equation}
where $\ell$ is the number of cuts in the cut system and $s$ is the number of connected
components $(\mathbb P_{\!i},\varGamma_{\!\!\!i})$ of the surface that is obtained from
$(E,\varGamma)$ upon cutting. This way $\widetilde\Bl_{E,\varGamma}$ is defined recursively 
by starting from spheres with at most three holes.
It can be proven ($\!$\cite[Sect.\,8.2]{Lyubashenko:Ribbon}, compare also \cite[Prop.\,3.4]{Fuchs:2016whb})
that the so defined functor can be concretely expressed as
  \begin{equation}
  \widetilde\Bl_{E,\varGamma}(-) \,\cong\, \Hom_\D({\bf 1}\,,-\otimes K^{\otimes g})\,.
  \end{equation}

In the second step of the construction, we make use of a right Kan extension $\mathrm R_U$
along the unmarking functor 
$U \colon \FM(E) \,{\xrightarrow{\,\simeq\,}}\, E/\!/ \mathrm{Map}(E)$ to obtain from
$\widetilde\Bl$ the left exact functor $\Bl$ that is part of the modular functor:
  \begin{equation}
  \begin{tikzcd}[row sep=3.1em, column sep=3.7em]
  \FM(E) \ar{r}{\widetilde\Bl} \ar{d}[swap]{U}
  & \mathcal{L}\!\text{\it ex}(\D^{\boxtimes p},\D^{\boxtimes q})
  \\
  E/\!/ \mathrm{Map}(E) \ar[dashed]{ur}[swap,xshift=-1pt]{\Bl}
  \end{tikzcd}
  \end{equation}
It can now be shown \cite{Fuchs:2016wjr}:

\begin{theo}
The right Kan extension $\Bl \,{:=}\, \mathrm R_U(\widetilde\Bl)$ exists and has a natural
monoidal structure.
\end{theo}


\section{Bulk correlators for non-semisimple conformal field theories} \label{seccs:3}

\subsection{Pinned block functors}

The modular functor developed above provides a framework that allows us
to address the issue of describing bulk fields and finding their correlators in full
non-se\-mi\-sim\-ple conformal field theories.
(This covers in particular the finitely semisimple case, as well as a large class of
logarithmic conformal field theories).

A first input datum of this description is an object $F\,{\in}\,\D$, to which we refer 
as the \emph{bulk object} of a full conformal field theory based on $\D$. To understand
the significance of this object, recall the description of bulk fields in
the semisimple case, that is, for rational conformal field theory. In that case
the chiral data form a finitely semisimple modular category
$\C$. Bulk fields are obtained by `combining left movers and right movers', whereby
they form an object $F$ in the enveloping category 
  \begin{equation}
  \C\,{\boxtimes}\, \C^\text{rev} =: \,\D \,.
  \end{equation}
By semisimplicity, $F$ can then be decomposed into a direct sum
  \begin{equation}
  F \,\cong \bigoplus_{i,j\in\pi_0^{}(\C)}Z_{i,j} \, S_i \boxtimes S_j \,,
  \end{equation}
with pairwise non-isomorphic simple objects $S_i$ of $\C$ and multiplicities
$Z_{i,j} \,{\in}\, \mathbb{Z}_{\geq 0}$. The partition function of bulk fields (that
is, the zero-point correlator of the full CFT on the torus)
is then a corresponding sesquilinear combination of the characters of the objects $S_i$,
with the same multiplicities $Z_{i,j}$.

An object of $\C\,{\boxtimes}\, \C^\text{rev}$ of the form $X\,{\boxtimes}\,Y$ 
is called $\boxtimes$-fac\-torized. In a non-semisimple conformal field theory, the
bulk object $F \,{\in}\, \D$ will no longer be a direct sum of $\boxtimes$-fac\-to\-rized
objects or, in other words, there is no longer a simple splitting of bulk fields
into left movers and right movers.
Accordingly, for now we select an arbitrary object $F$ of $\D$ as a candidate
bulk object $F$. Not surprisingly, to actually describe bulk fields this object will have
to be endowed with further structure (which we will exhibit below).
We then use the modular functor $\Bl$ to construct another functor, which we
denote by $\Bl^{(F)}$ and call the \emph{pinned block functor}, by evaluating $\Bl$
on the object $F$ in each of its arguments.

As in the case of $\Bl$, we start by first considering a functor $\widetilde\Bl^{(F)}\!$ on
surfaces with marking. Since correlators have to be compatible with sewing, it 
is convenient to include sewings of surfaces as additional
non-invertible morphisms. Thus we consider a category $\mSurfc$ with objects being finely
marked surfaces and morphisms being combinations of admissible moves and sewings.
(A move is called admissible if it corresponds to an element of the
mapping class group $\mathrm{Map}(E)$ as defined above, rather than to the larger 
group of mapping classes that do not necessarily preserve the orientation of 
every boundary circle. Furthermore, one must in fact consider a central extension
of the mapping class group; we suppress this aspect in our present brief exposition
and refer for details to Section 3.2 of \cite{Fuchs:2016wjr}.)

We are then in a position to construct the pinned block functor as a functor
  \begin{equation}
  \widetilde\Bl^{(F)}:\quad \mSurfc\, \xrightarrow{~~} \mathrm{vect} \,.
  \end{equation}
On objects, we define it by 
  \begin{equation}
  \widetilde\Bl^{(F)}(E,\varGamma) \,:=\, \widetilde\Bl_{E,\varGamma}(F,F,...\,,F) \,.
  \end{equation}
with $\widetilde\Bl_{E,\varGamma}$ the functors constructed in Section \ref{sec32}.

On moves, the functor yields linear maps that are defined in terms of 
linear isomorphisms for the elementary moves, explicitly expressed through
algebraic structure in $\D$. For example, for $X,Y\,{\in}\,\D$ the $Z$-move is mapped to
the linear map from $\Hom_\D(\mathbf1,X\,{\otimes}\, Y)$
to $\Hom_\D(\mathbf1,Y\,{\otimes}\, X)$ that acts as
  \begin{equation}
  f \,\longmapsto\, (d_X \,{\otimes}\, \mathrm{id}_{Y\otimes X}) \circ
  (\mathrm{id}_{X^\vee_{\phantom|}} \,{\otimes} f \,{\otimes}\, \pi_X^{-1})
  \circ b_{X^\vee_{\phantom|}} 
  \end{equation}
with $d$ and $b$ the evaluation and coevaluation of the right duality of $\D$
and $\pi$ the canonical pivotal structure of the ribbon category $\D$; pictorially,
  \begin{equation}
  \begin{picture}(240,73)(0,0)
  \put(46,7)      {\scalebox{.38}{\includegraphics{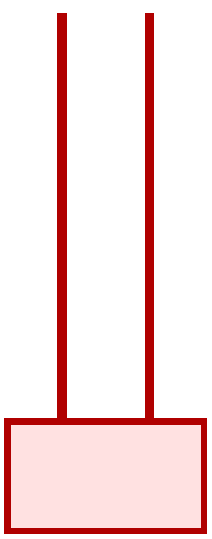}} }
  \put(49.5,68)   {$\scriptstyle X$}
  \put(55,12.1)   {$\scriptstyle f$}
  \put(60.1,68)   {$\scriptstyle Y$}
  \put(81,32)     {\large{\boldmath{$ \xmapsto{~\text{{\it Z}-move}~}$}}}
  \put(134,0)     {\scalebox{.38}{\includegraphics{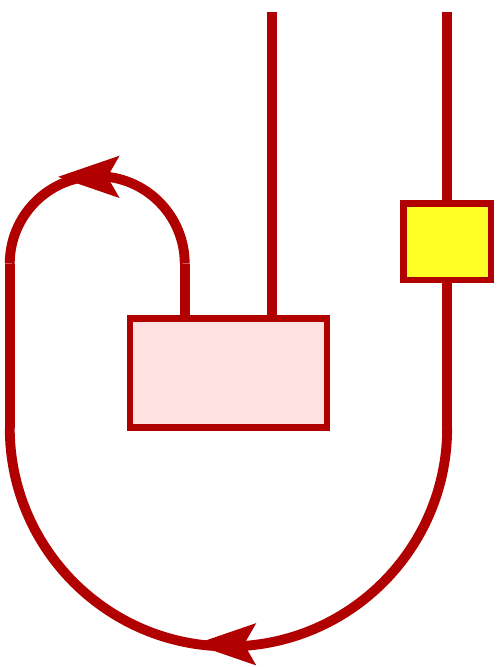}} }
  \put(156,30.6)  {$\scriptstyle f$}
  \put(161.4,75.5){$\scriptstyle Y$}
  \put(180.5,75.5){$\scriptstyle X$}
  \end{picture}
  \end{equation}
(Notice that this map needs to be defined for arbitrary objects $X,Y\,{\in}\,\D$, rather than
only for $X\,{=}\,F\,{=}\,Y$, because objects other than $F$ still occur as intermediate states.)
Finally, for realizing the sewing we use the structure morphism 
$\iota_F^K\colon F^\vee\,{\otimes}\, F \,{\to}\, K$ of the coend $K$, e.g.\
  \begin{equation}
  \Hom_\D(\mathbf1,K^{\otimes g}{\otimes}\, F^\vee{\otimes}\, F)\xrightarrow{~(\iota_F^K)_*^{}}
  \Hom_\D(\mathbf1,K^{\otimes(g+1)})
  \end{equation}
describes the sewing of a genus-$g$ surface with two punctures to a
genus-$(g{+}1)$ surface without any punctures.

\medskip

One then shows \cite[Prop.\,3.14]{Fuchs:2016wjr}:

\begin{propo}
This assignments of linear maps respect all 13 types of relations among the elementary
moves as well as the relations among sewings, and among moves and sewings. Thus it 
defines a functor $\widetilde\Bl^{(F)}\!\colon \mSurfc\,{\to}\,\mathrm{vect}$.
\end{propo}

Further, let $\Surfc\,\,$ be the category whose objects are extended surfaces and whose 
invertible morphisms are elements of $\mathrm{Map}(E)$ and which has further 
non-in\-ver\-ti\-ble morphisms given by sewings, as well as combinations of mapping class
group elements and sewings. Then one can consider a similar Kan extension
\begin{equation}
  \begin{tikzcd}[row sep=3.1em, column sep=3.7em]
  \mSurfc \ar{r}{\widetilde\Bl^{(F)}} \ar{d}[swap]{U} & \mathrm{vect}
  \\
  \Surfc \ar[dashed]{ur}[swap,xshift=-3pt]{\Bl^{(F)}}
  \end{tikzcd}
  \end{equation}
as for the modular functor $\Bl \,{=}\, \mathrm R_U(\widetilde\Bl)$ in Section \ref{sec32}
and finds \cite[Prop.\,3.15]{Fuchs:2016wjr}:

\begin{propo}
The right Kan extension $\mathrm R_U(\widetilde\Bl^{(F)}) \,{=:}\,\Bl^{(F)}$ of 
$\widetilde\Bl^{(F)}$ along the unmarking functor $U\colon \mSurfc\,{\to}\, \Surfc$ 
(defined analogously as the unmarking functor in the case of $\widetilde\Bl$)
exists and has a natural symmetric monoidal structure.
\end{propo}


\subsection{Consistent systems of correlators} \label{seccs:4a}

As a final ingredient we introduce a \emph{functor of trivial blocks}:
this is the functor $\bigtriangleup\colon \Surfc \,{\to}\, \mathrm{vect}$ that assigns
the ground field to any surface and the identity map on the ground field to any morphism
in $\Surfc$. (Likewise, we denote by $\widetilde\bigtriangleup\colon \mSurfc \,{\to}\, \mathrm{vect}$
the analogous functor on marked surfaces.) We can then
give a convenient characterization of a consistent system of correlators:

\begin{defi}
Let $\D$ be a modular finite ribbon category and $F\,{\in}\, \D$ an object.
A {\it consistent system $v_F^{}$ of bulk field correlators} with bulk object $F$ is a 
monoidal natural transformation
  \begin{equation}
  \begin{tikzcd}[column sep=3.7em]
  \Surfc~~~~ \ar[out=50,in=130]{r}[name=Arrow,below]{}
             \ar[out=50,in=130]{r}{~~~~\bigtriangleup}
             \ar[out=-50,in=-130]{r}[name=Brrow,above]{}
             \ar[out=-50,in=-130]{r}[swap,yshift=3pt]{~~\Bl^{(F)}}
  & ~\mathrm{vect}
  \ar[Rightarrow,to path= (Arrow) -- (Brrow) \tikztonodes]{}[swap]{v_F^{}}
  \end{tikzcd}
  \end{equation}
such that the morphism $v_F(E_{1|1}^0) \,{\in}\, \mathrm{End}_\D(F)$ is invertible. 
\\
(The latter condition is sufficient to normalize the bulk correlators and
excludes the trivial solution for which every component of the natural transformation
is the zero morphism.)
\end{defi}

It should be appreciated that this definition indeed encodes the covariance 
of correlators under sewing and their invariance under the action of
the relevant mapping class group. This holds simply because the very definition
of a natural transformation amounts to having commuting diagrams
  \begin{equation}
  \begin{tikzcd}[row sep=2.8em, column sep=3.3em]
  \Bbbk \ar{r}{\mathrm{id}_\Bbbk} \ar{d}[swap,yshift=3pt]{(v_F^{})_E^{}}
  & \Bbbk \ar{d}[yshift=2pt]{(v_F^{})_{E'}^{}}
  \\
  \Bl^{(F)}(E) \ar{r}{} & \Bl^{(F)}(E')
  \end{tikzcd}
  \end{equation}
in which $\Bbbk$ is the ground field (that is, the complex numbers in the 
application to CFT) and the arrow in the bottom row is any combination of the action 
of an element of $\mathrm{Map}(E)$ and a sewing.

The strategy of our construction of the natural transformation $v_F$ is similar
to what has been done in the construction of the modular functor: 
In a first step we construct a monoidal natural transformation 
  \begin{equation}
  \begin{tikzcd}[column sep=3.7em]
  \mSurfc~~~~~~~ \ar[out=50,in=130]{r}[xshift=6pt]{\widetilde\bigtriangleup}[name=Arrow,below]{}
  \ar[out=-50,in=-130]{r}[xshift=16pt,yshift=-10pt]{\widetilde{\Bl}^{(F)}}[name=Brrow,above]{}
  & ~\mathrm{vect}
  \ar[Rightarrow,to path= (Arrow) -- (Brrow) \tikztonodes]{}[swap]{\widetilde v_F^{}}
  \end{tikzcd}
  \end{equation}
that is analogous to $v_F$ but involves marked surfaces instead of surfaces. We refer to
this natural transformation $\widetilde v_F^{}$ as a system of \emph{pre-correlators}.
In a second step we combine the right Kan extension 
$\Bl^{(F)} \,{=}\, \mathrm R_U(\widetilde\Bl^{(F)})$ of $\widetilde\Bl^{(F)}$
with the trivial right Kan extension 
  \begin{equation}
  \begin{tikzcd}[row sep=2.9em]
  \mSurfc \ar{rr}{\widetilde\bigtriangleup}[name=Arrow,below]{} \ar{d}[swap]{U}
  & ~ & \mathrm{vect}
  \\
  \Surfc
  \ar[out=10,in=230,dashed,yshift=-2pt]{urr}[swap,xshift=-9pt] {\bigtriangleup=R_U(\widetilde\bigtriangleup)}
  \ar[Rightarrow,xshift=-2pt,to path = -- (Arrow) \tikztonodes]{}[description]{\mathrm{id}}
  \end{tikzcd}
  \end{equation}
of trivial functors to obtain the diagram
  \begin{equation}
  \begin{tikzcd}[row sep=2.9em] 
  \mSurfc \ar{rr}{\widetilde\Bl^{(F)}}[name=Arrow,below]{} \ar{d}[swap]{U}
  & ~ & \mathrm{vect}
  \\
  \Surfc \ar[in=250,out=-5]{urr}[xshift=18pt,yshift=-2pt]{\bigtriangleup}
  \ar[Rightarrow,xshift=5pt,to path = -- (Arrow) \tikztonodes]{}[description]{\widetilde v_F^{}}
  \end{tikzcd}
  \end{equation}
and then use the universal property of the Kan extension to decompose $\widetilde v_F^{}$
uniquely into $v_F^{}$ and another natural transformation $\psi$:
  \begin{equation}
  \begin{tikzcd}[row sep=3.5em] 
  \mSurfc \ar{rrr}{\widetilde\Bl^{(F)}}[name=endNatTrafb,below]{} \ar[out=-70,in=166]{dr}[swap]{U}
  & ~ & ~ & \mathrm{vect}
  \\
  ~& \Surfc~
  \ar[out=46,in=200,xshift=3pt,yshift=-1pt]{urr}[name=endCorr,below]{}[xshift=21pt,yshift=4pt]{\Bl^{(F)}}
  \ar[in=270,out=357]{urr}[name=beginCorr]{}[swap,xshift=-12pt,yshift=21pt]{v_F^{}~~~~~~~~~\bigtriangleup}
  \ar[Rightarrow,xshift=-5pt,to path=-- (endNatTrafb) \tikztonodes]{}[description]{\psi}
  \ar[Rightarrow,to path=(beginCorr) -- (endCorr)] {}
  \end{tikzcd} 
  \end{equation}

Besides the object $F$, the construction of $\widetilde\Bl^{(F)}$ uses as a basic input 
three morphisms $v_{0|3}^0 \,{\in}\, \Hom_\D(\mathbf1\,, F \,{\otimes}\, F \,{\otimes}\, F)$, 
$v_{1|0}^0 \,{\in}\, \Hom_\D(F\,,\mathbf1)$ and $v_{1|1}^0 \,{\in}\, \Hom_\D(F\,,F)$,
which play the role of candidates for the correlators
for the surfaces $ E_{0|3}^0$, $E_{1|0}^0$ and $ E_{1|1}^0$, i.e.\ for the
correlators on the sphere of three outgoing bulk fields, of one incoming bulk field,
and of a pair of an incoming and an outgoing bulk field, respectively.
We represent these morphisms pictorially as
  \begin{equation}
  \begin{picture}(220,36)(0,0)
  \put(-4,17)   {$ v_{0|3}^0 ~= $}
  \put(32,0)    {\scalebox{.27}{\includegraphics{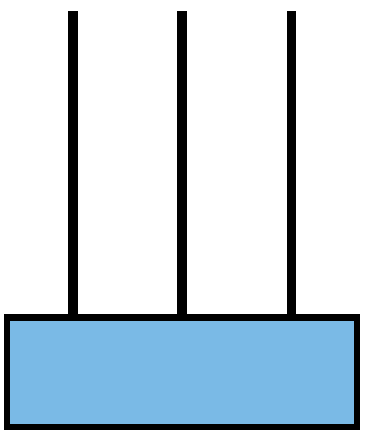}}}
  \put(94,17)   {$ v_{1|0}^0 ~= $}
  \put(130,9)   {\scalebox{.27}{\includegraphics{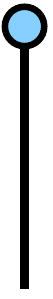}}}
  \put(169,17)  {$ v_{1|1}^0 ~= $}
  \put(205,0)   {\scalebox{.27}{\includegraphics{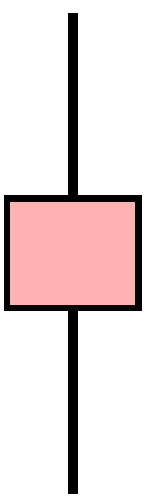}}}
  \end{picture}
  \end{equation}
Out of these morphisms we form further morphisms $\eta_F  ~{\in} ~ \Hom_\D(\mathbf1,F)$,~
$\mu_F  ~{\in} ~ \Hom_\D(F\,{\otimes}\,F,F)$,~ $\varepsilon_F ~{\in} ~ \Hom_\D(F,\mathbf1)$ 
 \\
and $\Delta_F\,{\in}\, \Hom_\D(F,F\,{\otimes}\,F)$
as follows: we set $\varepsilon_F \,{=}\, v_{1|0}^0$
and $\eta_F\,{:=}\,(\varepsilon_F \,{\otimes}\, \varepsilon_F \,{\otimes}\, \mathrm{id}_F)
\,{\circ}\, v_{0|3}^0$, while the other two morphisms are given pictorially as
  \begin{equation}
 \kern.4cm \begin{picture}(200,58)
  \put(-10,25)  {$ \mu_F  ~:= $}
  \put(25,-6)   {\scalebox{.27}{\includegraphics{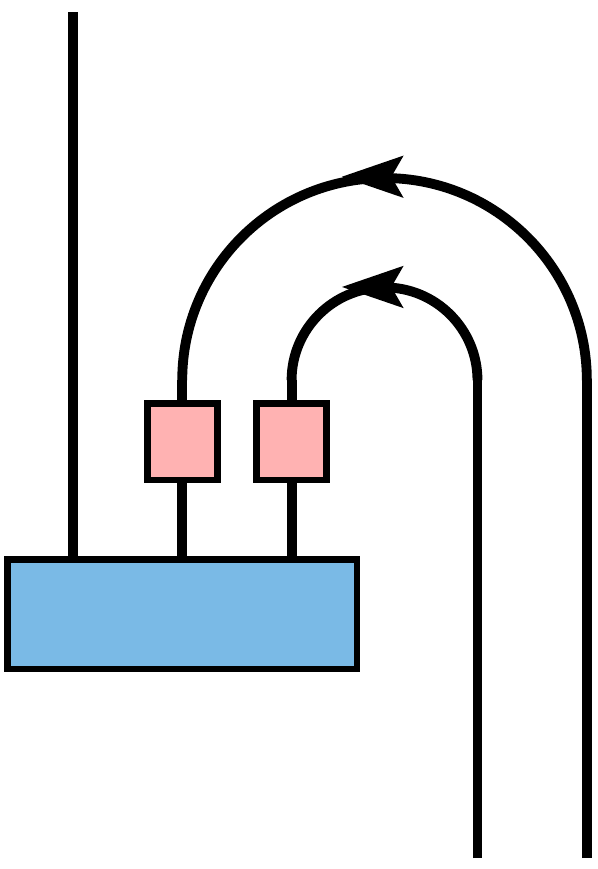}}}
  \put(95,25)   {and}
  \put(130,25)  {$ \Delta_F ~:= $}
  \put(165,-1)  {\scalebox{.27}{\includegraphics{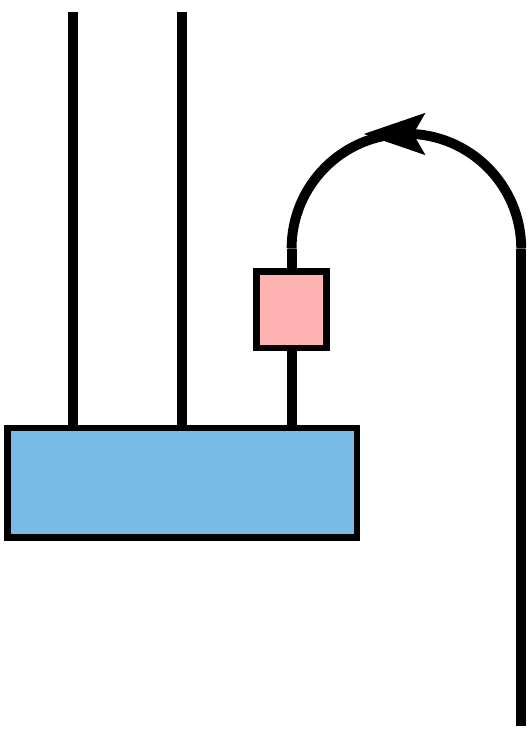}}}
  \end{picture}
  \end{equation}
These are candidate morphisms for endowing the object  $F$ with the structure of an algebra 
$(F,\mu_F,\eta_F)$ and of a coalgebra $(F,\Delta_F,\varepsilon_F)$.

To state our classification result for consistent sets of correlators, we further
introduce the following notion: A (co-)commutative symmetric Frobenius algebra 
$X \,{=}\, (X,\mu,\eta,\Delta,\varepsilon)$ in $\D$ is called \emph{modular} iff the equality
  \begin{equation}
  S^K \circ\, [ \,\iota^K_X \circ (\mathrm{id}_X \otimes \varPhi_X) \circ \Delta \,]
  \,=\, \iota^K_X \circ (\mathrm{id}_X \otimes \varPhi_X) \circ \Delta \,,
  \end{equation}
of morphisms in $\Hom_\D(X,K)$ holds. Here, as above, $K\,{\in}\, \D$ is the coend 
$\int^{Y\in\D}\! Y^\vee{\otimes}\,Y$, with structure morphism
$\iota^K_Y \,{\in}\, \Hom_D(Y^\vee{\otimes}\,Y,K)$, while
$S^K \,{\in}\, \mathrm{End}_\D(K)$ is an automorphisms
that, via post-composition, realizes the $S$-move on morphisms,
and $\varPhi_X \,{:=}\, ((\varepsilon \,{\circ}\, \mu) \,{\otimes}\, \mathrm{id}_{X^\vee_{}})
\,{\circ}\, (\mathrm{id}_X \,{\otimes}\, b_X)$ is an isomorphism in $\Hom_\D(X\,,X^\vee)$.

\begin{theo}\cite{Fuchs:2016wjr}
The input data $F$ and $v_{0|3}^0$, $v_{1|0}^0$, $v_{1|1}^0$ determine a consistent set
$v_F\colon \Delta\,\Rightarrow\, \Bl^{(F)}$ of bulk field correlators 
if and only if $(F,\mu_F,\Delta_F,\epsilon_F,\eta_F)$ is a modular Frobenius algebra.
\end{theo}

This result has a natural interpretation within the so-called \emph{microcosm
principle} which states that in order to define an algebraic structure in some categorical
framework, the category needs to be an object in a bicategory that has similar properties
when regarding it in the bicategorical setting. For instance,
monoids can only be defined in monoidal categories, and modules over
a monoid only in module categories over a monoidal category. And in the present context,
we have to deal with categories with dualities, and these are indeed Frobenius pseudo-monoids 
in the bicategory of categories \cite{Street:Frobenius}.

For the theorem above to be of relevance, we have to make sure that modular Frobenius algebras
exist. This has indeed been established for the case that the modular tensor category $\D$
is the category of finite-di\-men\-si\-onal modules over a finite-dimensional factorizable ribbon 
Hopf algebra:

\begin{theo}\cite{Fuchs:2012wh}
Let $H$ be a finite-dimensional factorizable ribbon Hopf algebra over an algebraically closed
field and $\omega\colon H \,{\to}\, H$ be a ribbon automorphism. Then
  \begin{equation}
  \int^{m\in H\text{-}\mathrm{mod}}\! \omega(m)^{\!\vee}\boxtimes m ~~\in\, H\text{-}\mathrm{bimod}
  \end{equation}
is a modular Frobenius algebra in the modular category $H\text{-}\mathrm{bimod}$.
\end{theo}

In particular, taking $\omega$ to be the identity morphism shows that the co-regular bimodule 
$H^*$ is a modular Fro\-be\-nius algebra. One expects that analogously the coend
  \begin{equation}
  F_{\!\circ} \,:= \int^{X\in\C}\!\! X^\vee\boxtimes X ~~ \in \, \C\,{\boxtimes}\, \C^\text{rev}
  \end{equation}
is a modular Frobenius algebra, not only when $\C$ is the representation category of a
factorizable ribbon Hopf algebra, but for any modular tensor category. The full CFT with
this coend as the bulk object is often called the \emph{Cardy case}.

It is worth pointing out that the derivation of the result that full conformal field theories are in
bijection with modular Frobenius algebras is constructive. In particular, as a by-product
it yields a universal formula for the bulk field correlators of any full CFT:
  \begin{equation}
  v_F^{}(E_{p|q}^g) \,=\, \Delta_F^{(q-1)}\circ\, \sigma_{\!F,K}^{(g)}\,\circ\, \mu_F^{(p-1)} .
  \end{equation}
Here the symbol $\mu_F^{(\ell)}$ stands for any iterated product of $\ell{+}1$ factors of $F$,
$\Delta_F^{(\ell)}$ for any iterated coproduct, and
$\sigma_{F,K}^{(\ell)} \,{\in}\, \Hom_\D(F,F\,{\otimes}\,K^{\otimes \ell})$ for an iteration
of the morphism
  \begin{equation}
  \sigma_{\!F,K}^{} := (\mu_F \,{\otimes}\, \iota_F^K) \circ
  (\mathrm{id}_F \,{\otimes}\, b_F \,{\otimes}\, \mathrm{id}_F) \circ \Delta_F
  \, \in \Hom_\D(F,F\,{\otimes}\,K) \,.
  \end{equation}
  Graphically, the morphisms $\sigma_{\!F,K}^{(\ell)}$ are given by
  \begin{equation}
  ~~~~\begin{picture}(220,90)
	  \put(0,-5){
  \put(-7,46)       {$ \sigma_{\!F,K}^{(1)} ~=~ \sigma_{\!F,K}^{} ~= $}
  \put(65,11)       {\scalebox{.19}{\includegraphics{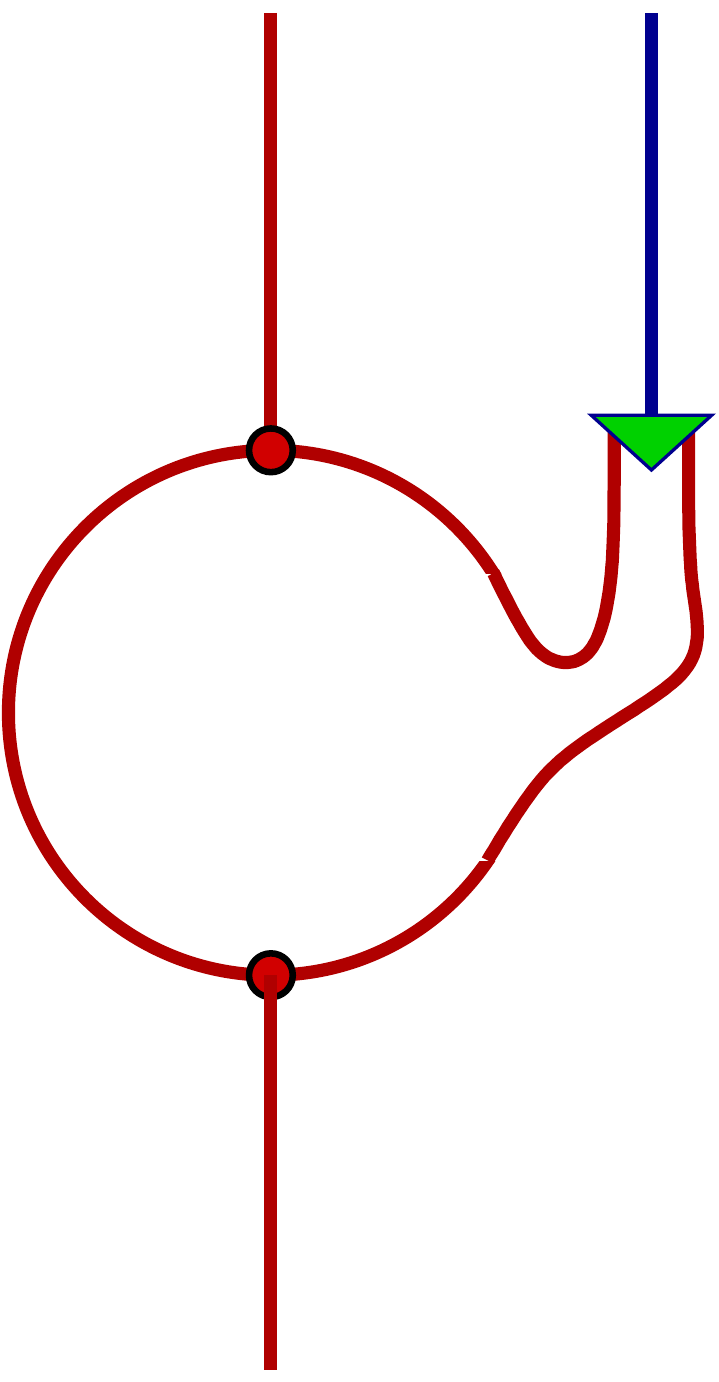}}}
  \put(76.5,2.4)    {$\scriptstyle F$}
  \put(76.9,89.3)   {$\scriptstyle F$}
  \put(81.1,26.9)   {$\scriptstyle \Delta_F$}
  \put(81.0,64.8)   {$\scriptstyle \mu_F$}
  \put(98.0,89.3)   {$\scriptstyle K$}
  \put(104.7,61.7)  {$\scriptstyle \iota_F^K$}
  \put(138,46)      {$ \sigma_{\!F,K}^{(2)} ~=~ $}
  \put(177,7)       {\scalebox{.19}{\includegraphics{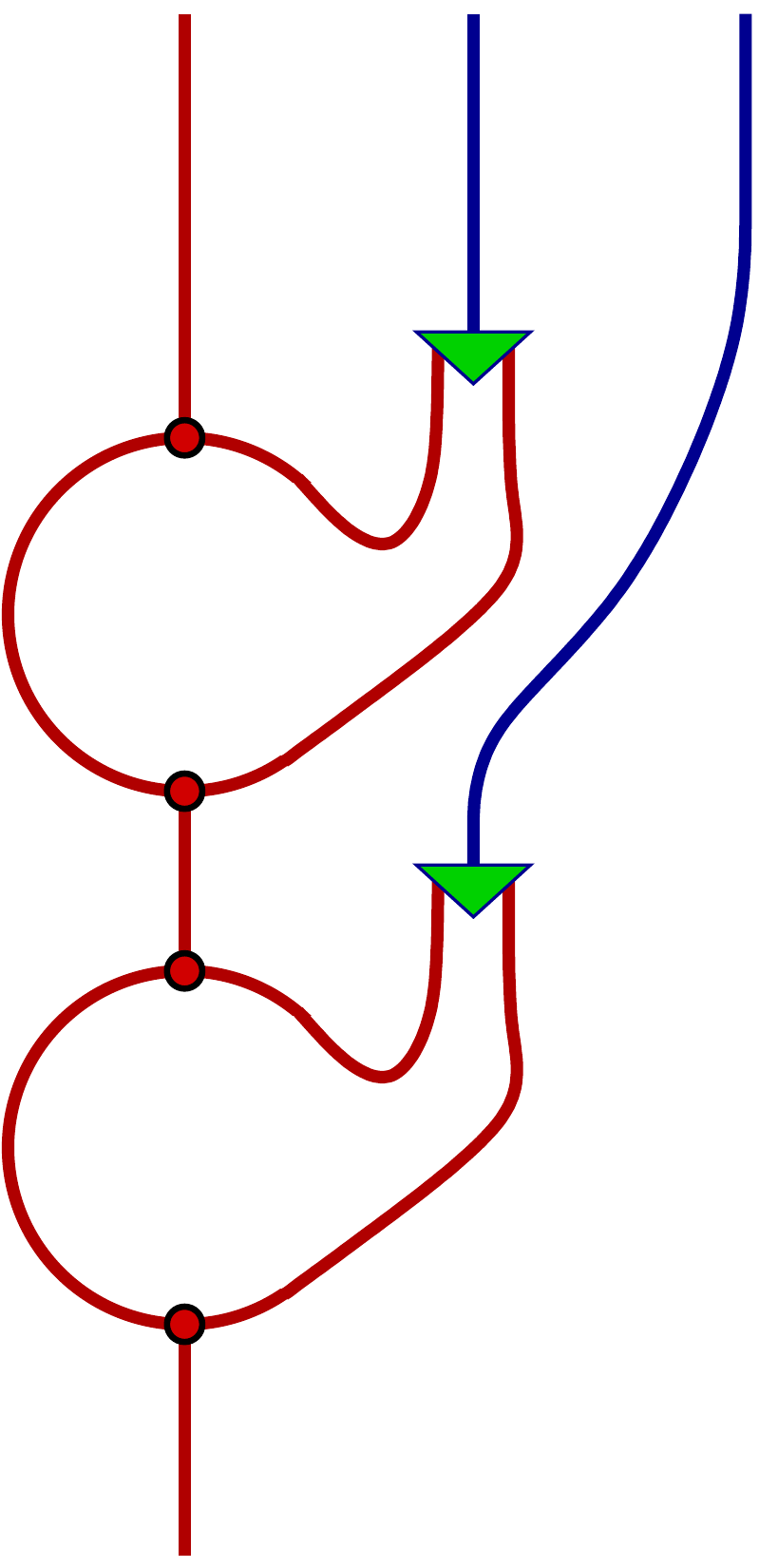}}}
  \put(184.5,-0.6)  {$\scriptstyle F$}
  \put(184.8,100.3) {$\scriptstyle F$}
  \put(201.2,100.3) {$\scriptstyle K$}
  \put(217.1,100.3) {$\scriptstyle K$}
  }
  \end{picture}
  \end{equation}
etc. Thus a graphical description of the correlator is
  \begin{equation}
  \kern-7pt\begin{picture}(140,143)
  \put(7,63)     {$ v_F^{}(E_{p|q}^g) ~= $}
  \put(50,124.2) {$\scriptstyle \Delta_F^{\!(q-1)}$}
  \put(69,-5)    {\scalebox{.19}{\includegraphics{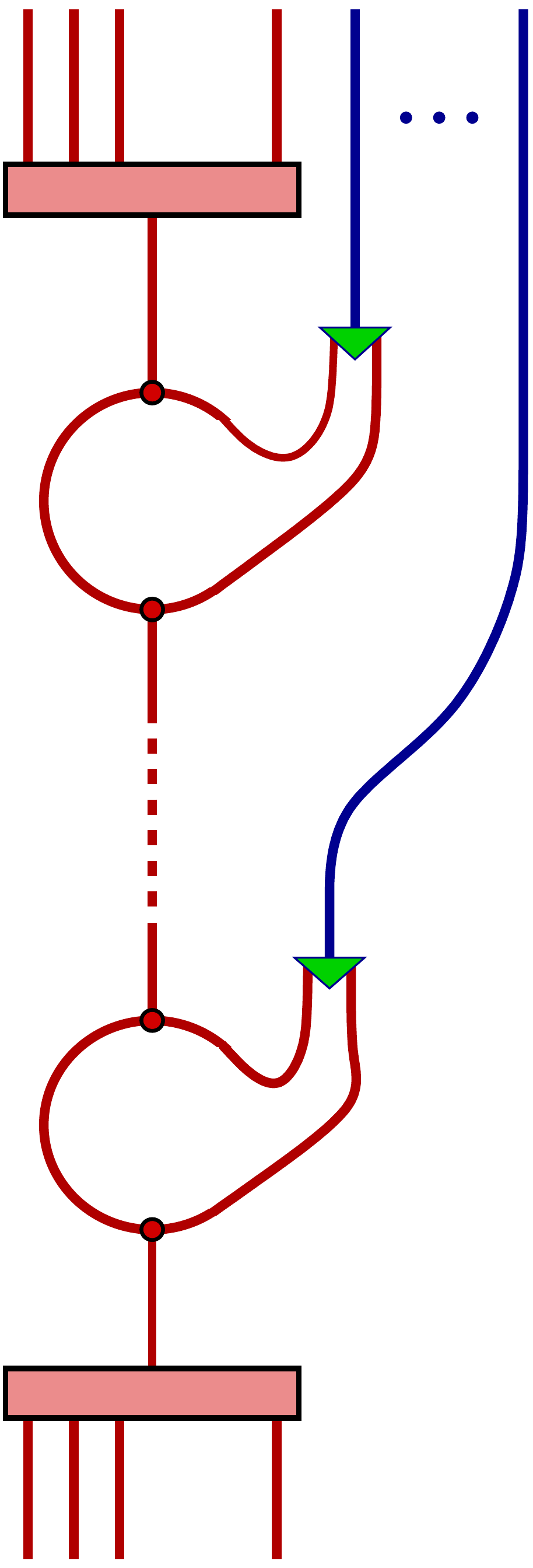}}}
  \put(81,-6)    {$\scriptstyle F^{\otimes p}_{}$}
  \put(81.5,139) {$\scriptstyle F^{\otimes q}_{}$}
  \put(98.6,10.1){$\scriptstyle \mu_F^{(p-1)}$}
  \put(103.7,139){$\scriptstyle K^{\otimes g}_{}$}
  \end{picture}
  \end{equation}
(A priori, the formula realized by this picture is the one obtained for the \emph{pre}-correlator 
$\widetilde v_F^{}(E_{p|q}^g,\varGamma)$ for a specific fine marking $\varGamma$ on the
surface $E_{p|q}^g$. However, any other choice of $\varGamma$ yields an expression that, by the 
various properties of the structure morphisms of $F$ and $K$, gives exactly the same morphism.
For instance, the factors in the iterated product may be multiplied in any
arbitrary order, owing to the associativity and commutativity of the multiplication $\mu_F$.)

The closed formula for $v_F^{}(E_{p|q}^g)$ may be suggestively interpreted as resulting from the
following prescription:

\smallskip

\begin{enumerate}[i)]
\item
Draw a skeleton for the surface $E_{p|q}^g$ that has trivalent vertices and includes an
outward-oriented `external' edge attached to every boundary component in such a way that the subgraph 
formed by the external edges for all outgoing boundary components is a connected tree, and likewise 
for the one formed by the edges for all incoming boundary components, and such that each loop
of the graph consists of precisely two edges. Label every edge of this
skeleton with the Frobenius algebra $F$ in $\D$.
\\
(Instead of considering a graph, a priori one may want to work with a ribbon graph. But this is
insignificant because, as a consequence of its other properties, $F$ also has trivial twist, 
$\theta_F \,{=}\, \mathrm{id}_F$.)

\item
Orient the internal edges of the skeleton in such a way that each of the vertices 
either has one outgoing and two incoming edges or vice versa. In the former case,
label the vertex with the product $\mu_F$, and in the latter case with the coproduct
$\Delta_F$ of $F$.

\item
Further, for each of the $g$ handles of the surface $E_{p|q}^g$, attach an additional edge to
one of the two edges of the corresponding loop of the skeleton. 
\\
Label these edges by the Hopf algebra $K \,{\in}\, \D$, and label the resulting new trivalent
vertices by the component $\iota_F^K$ of the dinatural transformation that comes with
the coend $K$.

\item
Interpret the so obtained graph as a morphism in $\D$. 
\end{enumerate}


\section{The Cardy--Cartan partition function} \label{seccs:4b}

Historically, the description of the bulk fields of a full conformal field theory has been 
intimately linked to the classification of modular invariants -- that is, in the fi\-ni\-te\-ly
semisimple case, of square matrices $\big(Z_{i,j}\big)_{i,j\in\pi_0(\C)}^{}$ with non-negative
integral entries and with $Z_{0,0} \,{=}\, 1$ that commute with the
representation of the modular group SL$(2,\mathbb Z)$ on the characters of the chiral CFT.
The simplest such modular invariant is the charge conjugation invariant
$Z_{i,j} \,{=}\, \delta_{i,j^\vee}$, which for finitely semisimple $\C$ corresponds to the Cardy case.

We now give a brief description of the modular invariant torus partition function for the Cardy case
when $\C$ is not semisimple, restricting to the case that 
$\C \,{\simeq}\, H\text{-mod}$ is the category of modules over a (non-semisimple)
finite-dimensional factorizable ribbon Hopf algebra $H$.

We first recall that for any algebra $A$ in a pivotal category $\C$ and any left $A$-module
$(M,\rho)$ with $A$-action $\rho\colon A\,{\otimes}\, M\to M$, taking a partial trace of the
representation morphism $\rho$ gives the character $\chi_M^A \,{\in}\,\Hom_\C(A,\mathbf1)$.
Further the modular Frobenius algebra $F_{\!\circ}$ in the category
 \\[-1.8em]~
  \begin{equation}
  \begin{array}{ll}
  H\text{-bimod} ~~& \simeq\,\, H\text{-mod} \,\boxtimes\, H\text{-mod}^\text{rev}
  \\[2pt] &
  \simeq\,\, H\text{-mod} \,\boxtimes\, \text{mod-}H
  \end{array}
  \end{equation}
is the $\Bbbk$-vector space $H^*_{}$ with co-regular $H$-action from the left and right. We can 
compute the partition function 
  \begin{equation}
  Z = \chi_{F_{\!\circ}}^{K} ~\,\in\, \Hom_{H\text{-bimod}}(K,\mathbf1)
  \end{equation}
in the following way. The $H$-bimodule structure of the coend $K$ is as follows. $K$ is the $\Bbbk$-vector
space $H^* {\otimes}\, H^*$ with co-adjoint actions from the left and right on the first and 
second factor, respectively. As a consequence, we can express the character $\chi_{F_{\!\circ}}^K$
as a linear combination of products of characters of left- and right-mo\-du\-les, respectively,
over the $H$-module Hopf algebra $L \,{=}\, H^*\!$ (with co-adjoint $H$-action) in $H$-mod. 
Moreover, the latter can be written
in terms of $\Bbbk$-algebra characters $\chi_M^H \,{\in}\,\Hom_\Bbbk(H,\Bbbk)$ via
  \begin{equation}
  \chi_M^{L} = \chi_M^H \circ \mu_H \circ (f_Q\otimes uv^{-1}) 
  \end{equation}
where $f_Q\colon H^* \,{\to}\, H$ is the Drinfeld map, $u$ the Drinfeld
element and $v$ the ribbon element of $H$. Then textbook results in algebra
can be applied to write $Z$ in the form \cite{Fuchs:2012ya}
  \begin{equation}
  Z = \sum_{i,j\in \pi_0(\C)} c_{i,j}^{}\; \chi^L_i \otimes \chi^L_j \,,
  \end{equation}
where $c$ is the \emph{Cartan matrix} of the category $H$-mod, i.e.\
  \begin{equation}
  c_{i,j}^{} = [P_j,S_i] = \dim_\Bbbk\Hom_{H\text{-mod}}(P_i,P_j)
  \end{equation}
with $P_i$ non-isomorphic indecomposable projective objects of $H$-mod, which are
the projective covers of the simple objects $S_i$.

The so obtained combination $Z$ of characters is not only modular invariant, but is,
by the previously summarized results, indeed also the torus partition function for
the consistent Cardy-case full conformal field theory. Accordingly we call it the
\emph{Cardy--Cartan partition function} for $\C \,{=}\, H\text{-mod}$.
Since the Cartan matrix $c$ is expressed in terms of categorical quantities,
without direct reference to $H$, it is natural to expect that the same formula
gives in fact the Cardy-case torus partition function for any finite tensor category $\C$.


\section{Results on boundary states} \label{seccs:4}

The field content of a local conformal field theory can be expected to be
much richer than merely consisting of bulk fields; apart from these, it also comprises
boundary fields as well as defect fields. A first step for extending the results of
Sections \ref{seccs:4a} and \ref{seccs:4b} to 
that general situation is a discussion of boundary fields and, more specifically, of the
corresponding \emph{boundary states}, which describe the one-point correlators of bulk
fields on a disk. 

The following three postulates appear to be natural:

\smallskip

\begin{itemize}
\item[({\bf BC})]
Boundary conditions are objects of a category.  
\\
$~$\hspace*{1.53em}(In the Cardy case, this category should be $\C$.)

\item[({\bf BS})]
A boundary state is an element of the \emph{center}
  \begin{equation}
  ~~~\begin{array}{ll}
  \mathrm{End}(\mathrm{Id}_\C) \! & \displaystyle
  = \int_{c\in\C} \Hom_\C(c,c) 
  \\{}\\[-10pt] & \displaystyle
  \cong \int_{c\in\C} \Hom_\C(c^\vee{\otimes}\, c,1) \,\cong\, \Hom_\C(L,\mathbf 1)
  \end{array}
  \end{equation}
$~$\hspace*{0.65em}of $\C$, where $L \,{=}\, \int^{c\in\C}\! c^\vee{\otimes}\, c$.

\item[({\bf F})]
In the Cardy case the algebra of bulk fields is given by 
\\
$~$\hspace*{0.78em}the object
  \begin{equation}
  ~~~~F_{\!\circ} = \int^{c\in \C}\!\!\! c^\vee\boxtimes c ~~ \in~ \C\boxtimes\C^\text{rev}
  \end{equation}
$~$\hspace*{0.65em}endowed with its canonical Frobenius structure.
\end{itemize}

\smallskip

A boundary state amounts to mapping the objects of $\C$ to $\mathrm{End}(\mathrm{Id}_\C)$, and thus
to a decategorification. It is therefore natural to expect that incoming and outgoing
boundary states factor through characters 
  \begin{equation}
  \chi^L_m ~\in \Hom_\C(L,\mathbf1) 
  \end{equation}
and through cocharacters
  \begin{equation}
  \widehat \chi^L_m ~\in \Hom_\C(\mathbf1,L) \,,
  \end{equation}
respectively, of the Hopf algebra $L \,{\in}\, \C$.

It turns out that when imposing the postulates ({\bf BC}), ({\bf BS}) and ({\bf F}),
boundary states are indeed consistently given by (co)characters up to composition with
the dinatural family for the coend $L$ in the case of outgoing boundary states, respectively
with the one for the \emph{end} $\,L^\vee$, which can be identified with $L$ via the non-degenerate
Hopf pairing. This leads in particular to

\begin{theo} \cite{Fuchs:2017unc} 
In the Cardy case, the sewing of the one-point correlators for bulk fields on two disks 
with boundary conditions $m,n \,{\in}\, \C$, respectively, results in an annulus
partition function $\mathrm A_{m,n}$ that expands as
  \begin{equation}
  ~~\mathrm A_{m,n}
  = \sum_{i\in \pi_0(\C)} \dim_\Bbbk \Hom_\C(m\,{\otimes n},S_i)\, \widehat \chi_i^L
  \end{equation}
in terms of simple $L$-cocharacters.
\end{theo}

In particular, also for chiral data that are non-semi\-sim\-ple, the annulus partition
functions can be written as an linear combination of characters with non-negative integral
coefficients, as befits a partition function. It is remarkable that precisely as in the
semisimple case the annulus coefficients are in the Cardy case given by the fusion rules.
This result is in fact particularly strong for categories $\C$ that are not semisimple, because
in that case the subspace of $\Hom_\C(L,\mathbf1)$ that is spanned by the characters
is a proper subspace.


\section{Towards derived modular functors} \label{seccs:5}

Recall from Section \ref{seccs:2} that to any, not necessarily semi\-sim\-ple,  
modular tensor category $\C$ there is associated a modular functor $\Bl$, which to a surface
$E_{p|q}^g$ assigns a left exact functor $\Bl_{E_{p|q}^g}\!\colon \C^{\boxtimes p} \,{\to}\, 
\C^{\boxtimes q}$. 

Let us for simplicity consider the case $q\,{=}\,0$, i.e.\ that all
boundary circles are incoming. Then we deal with a functor
$\Bl_{E_{p|0}^g}\colon \C^{\boxtimes p} \,{\to}\, \mathrm{vect}$ with
  \begin{equation}
  \Bl_{E_{p|0}^g} (X_1\,{\boxtimes}\,\cdots \,{\boxtimes}\, X_p)
  \,\cong\, \Hom_\C(X_1\,{\otimes}\,\cdots \,{\otimes}\, X_n,L^{\otimes g}) \,,
  \end{equation}
carrying an action of the mapping class group $\mathrm{Map}(E_{p|0}^g)$ 
by natural endotransformations and being compatible with sewing.

For non-semisimple $\C$ the Hom functor is only left exact and hence has
derived functors. This is not an artefact of our approach to conformal
blocks: Conformal blocks are invariants, and in general taking invariants is
not an exact functor. It is therefore natural to ask whether in the non-semisimple
case the mapping class groups act not only on Hom-spaces, but on $\mathrm{Ext}$-spaces as well.

Indeed, via a subtle interplay between the monoidal structure and homological algebra one shows:

\begin{theo}
\begin{enumerate}[i)]
\item
The mapping class group $\mathrm{Map}E_{p|0}^g$ naturally acts on the space
$\mathrm{Ext}^p_\C(X_1\,{\otimes}\,\cdots \,{\otimes}\, X_p,L^{\otimes g})$.

\item
In particular \cite{Lentner:1707.04032}, the modular group $\mathrm{SL}(2,\mathbb{Z})$ acts
(projectively) on the Hochschild complex of a factorizable ribbon Hopf algebra.
\end{enumerate}
\end{theo}

\noindent
We briefly sketch the idea of the underlying construction:

\smallskip

\begin{enumerate}[i)]
\item
Fix a surface $E_{p|0}^g$ of genus $g$ with $p$ disjoint boundary circles.
\item
Fix a projective resolution $P_{\!\bullet} \,{\to} \mathbf1$ of the
monoidal unit of $\C$ and insert it at an auxiliary circle in $E_{p|0}^g$
that is disjoint from the other $p$ boundary circles.
\item
The functoriality of $(p{+}1)$-point blocks
  \begin{equation}
  \mathrm{Bl}_{E_{p+1|0}^g}: \quad \C^{\otimes(n+1)} \to \mathrm{vect}
  \end{equation}
then gives a complex of left exact functors $\C^{\otimes n} \,{\to}\, \mathrm{vect} $ that
carries a (projective) action of the mapping class group $\mathrm{Map}(E_{p+1|0}^g)$ for
surfaces of genus $g$ with $p{+}1$ boundary circles.
\item
The kernel of the obvious surjection
  \begin{equation}
  \mathrm{Map}(E_{p+1|0}^g) \to \mathrm{Map}(E_{p|0}^g)
  \end{equation}
has an explicit description which can be used to show that it acts
trivially on the $\mathrm{Ext}$-vector spaces. As a consequence the action of
$\mathrm{Map}(E_{p+1|0}^g)$ descends to an action of $\mathrm{Map}(E_{p|0}^g)$
on the $\mathrm{Ext}$ spaces.
\end{enumerate}


\section{A few open questions} \label{seccs:6}

Various issues still need to be addressed before one can reach a complete understanding
of full local logarithmic conformal field theory. Major challenges are:

\smallskip

\begin{enumerate}[i)]
\item
Find a natural categorical description of the field content, covering in particular also
boundary fields and general defect fields.
\item
Describe the fundamental correlators for those fields (compare
\cite{Fuchs:2004xi} for the finitely semisimple case).
\item
Give a `holographic' construction of a full logarithmic conformal field 
theory from a $(2{+}\epsilon)$-dimensional to\-po\-logical field theory.
\\
(Such a construction, based on the three-dimensional topological field theory
of the Reshetikhin--Turaev sur\-ge\-ry construction is well established in the
finitely se\-mi\-sim\-ple case; see \cite{Schweigert:2006af} for a review.)
\end{enumerate}

\smallskip 

\noindent
We also mention the following conceptual questions:

\smallskip

\begin{enumerate}[i)]
\item
How stable are the results conceptually? 
\\
Specifically, are the main ideas
still applicable within more general (categorical) frameworks?
\item
In particular, how critical is the rigidity of the categories involved?
\item
Is there a role for logarithmic CFT -- or, more generally, for non-semisimplicity --
in string theory?
\\
(After all, bosonic ghosts can profitably be studied as a 
logarithmic conformal field theory \cite{Ridout:2014oca}.)
\item
Do `derived conformal blocks' have physical applications, 
e.g.\ in string theory or in statistical mechanics?
\end{enumerate}


\bibliography{allbibtex}

\bibliographystyle{prop2015}

\end{document}